\documentclass[a4paper,12pt]{article}
 \pdfoutput=1
\usepackage[]{inputenc,graphicx,amsmath,bbm,latexsym,textcomp,amssymb, mathrsfs,appendix,cite}
\usepackage{color, multicol,multirow}
\usepackage{subfigure,float, afterpage}  
\usepackage{hyperref}
\usepackage[margin=1.15in,top=1.1in]{geometry}

\newcommand{\beq}{\begin{equation}}
\newcommand{\eeq}{\end{equation}}
\newcommand{\bea}{\begin{eqnarray}}
\newcommand{\eea}{\end{eqnarray}}
\newcommand{\Z}{\mathbb{Z}}

\def\nn{\nonumber}

\def\ra{\rightarrow}
\def\mpl{M_{\rm Pl}}

\newcommand{\MS}{\overline{\mbox{\sc ms}}}

\def\etc{ {\em etc.\ }}
\def\ie{ {\em i.e.,\ }}

\begin{document}
 \begin{center}
{\Large{\bf  \color{black}
A study of electroweak vacuum metastability with a singlet scalar dark matter}}\\

\vspace {1.0cm}

{\bf Najimuddin Khan}$^\dagger$\footnote{phd11125102@iiti.ac.in} and
{\bf Subhendu Rakshit}$^\dagger$\footnote{rakshit@iiti.ac.in} \\
	
\vskip 0.15in
{\it
$^\dagger${Discipline of Physics, Indian Institute of Technology Indore,   \\
IET-DAVV Campus, Indore 452017, India  }\\
}
\vspace{0.40cm}
\end{center}
\begin{abstract}
 We study several aspects of electroweak vacuum metastability when an extra gauge singlet scalar, a viable candidate for a dark matter particle, is added to the standard model of particle physics, which is assumed to be valid up to the Planck scale. Phase diagrams are drawn for different parameter spaces,  and based on that, we graphically demonstrate how the confidence level, at which stability of electroweak vacuum is excluded, depends on such new physics parameters.
\end{abstract}

\renewcommand*{\thefootnote}{\fnsymbol{footnote}}

\newpage

\section{Introduction}
The recent discovery of the Higgs boson at $125.66\pm0.3$ GeV~\cite{Aad:2012tfa, Chatrchyan:2012ufa, Giardino:2013bma} completes search for the particle content of the standard model (SM), confirming Higgs mechanism to be responsible for electroweak symmetry breaking. It brings forward the issue of naturalness, indicating the presence of new physics at the $\sim $\,TeV scale.

LHC, with 20\,fb$^{-1}$ data in its $\sqrt s=8$ TeV run, is yet to find any such hint of new physics. Hence, it looks like infrared  naturalness~\cite{Giudice:2013yca} is not the right guiding principle for hunting  physics beyond SM, and, new ideas are required to understand hierarchy problem. It is possible that SM is not an effective theory but a fundamental theory of Nature, valid up to the Planck scale ($M_{\rm Pl}$), where quantum gravity effects kick off.

In such a scenario, Higgs boson mass $M_H\sim126$~GeV is consistent with $\lambda(M_{\rm Pl})\sim 0$, \ie with vanishing self-coupling of Higgs boson at $M_{\rm Pl}$. In addition, as the beta function
\beq
\beta_{M_H^2}=\frac{d\,M_H^2(\mu)}{d\,\ln\mu} \nn 
\eeq
is proportional to $M_H^2$, the smallness of the Higgs mass would be ensured if $M_H(M_{\rm Pl})\sim 0$~\cite{Giudice:2013yca, Holthausen:2013ota}. All these indicate a vanishing Higgs potential near $M_{\rm Pl}$, which perhaps is ensured by Planck scale physics. The exact mechanism for this is unknown, as we do not expect to understand the dynamics at this regime in the context of a conventional perturbative quantum field theory. So in the quantum corrections to the Higgs mass,  putting the cut-off scale $\Lambda\ra M_{\rm Pl}$  does not make any sense.
If $M_{\rm Pl}$ does not get introduced in the theory as a physical scale~\cite{Holthausen:2013ota},  there is no hierarchy problem to start with.  So, smallness of Higgs mass may rather be ensured by ultraviolet naturalness~\cite{Giudice:2013yca} than the usual infrared naturalness: The quadratic divergences might be spurious~\cite{Holthausen:2013ota, Bardeen:1995kv}, as we are looking from the electroweak scale than the Planck scale which possibly hold the key of ensuring naturalness.

For Higgs field values much greater than the electroweak (EW) vacuum, the Higgs effective potential~\cite{Coleman:1973jx, Sher:1988mj} turns  down due to large contributions from top quark loops. As a result, there might exist a second minimum of the potential around $\mpl$. If this minimum is significantly deeper than the EW minimum, the vacuum transition probability can be sizable, threatening the fate of this Universe.  Given the present uncertainties of SM parameters, the EW vacuum is metastable at 99.3\%~CL~\cite{Buttazzo:2013uya}.

Stability of the potential is ensured if the two vacua are at the same level, namely
\beq
V(\phi_{\rm EW})=V(\phi_{\rm new}) \quad {\rm and} \quad
V^\prime(\phi_{\rm EW})=V^\prime(\phi_{\rm new})=0 \nn \, ,
\eeq
where, $\phi_{\rm EW}$ and $\phi_{\rm new}$ correspond respectively to the EW vacuum and the new vacuum at a very large value of the field. Neglecting quantum corrections of $\cal{O}(\alpha)$, this corresponds to the conditions~\cite{Bezrukov:2012sa}
\beq
\lambda(\Lambda_0) = \beta_{\lambda}(\Lambda_0)=0 \, ,\qquad \qquad \beta_{\lambda}(\mu)=\frac{d\,\lambda(\mu)}{d\, \ln\mu}\, ,
\eeq
where $\Lambda_0$ is some energy scale in between the EW scale and $\mpl$, at which the instability starts setting in.

Recent cosmological and astrophysical evidences suggest presence of cold dark matter (DM), which cannot be explained in the scenario described above. A simple choice is to add a gauge singlet real scalar $S$ to the SM~\cite{Burgess:2000yq}. An additional $\Z_2$ symmetry ensures the stability of $S$. The scalar modifies the Higgs effective potential, and can ensure vacuum stability up to $\mpl$.

Such extensions of SM  have been discussed in the literature in the context of vacuum stability. In a model SM+$S$, where SM is extended by a scalar DM, whether the boundary conditions 
$\lambda=0$, $\beta_\lambda=0$ and the Veltman condition are satisfied at $\mpl$ is discussed 
in ref.~\cite{Haba:2013lga} (see also refs.~\cite{Kadastik:2011aa, Chen:2012faa, Gabrielli:2013hma}). It is shown that the correct DM relic density is obtained if DM mass lies in the range 300 GeV $\lesssim M_S \lesssim$ 1 TeV. 
Introduction of a scalar relaxes~\cite{Gonderinger:2009jp} Higgs mass bounds stemming from vacuum stability considerations and perturbativity. Such constraints were also considered in SM+$S$ model, in the context of inflation, where either Higgs~\cite{Clark:2009dc} or the scalar $S$~\cite{Lerner:2009xg} acts as an inflaton. Model dependence of Higgs mass bounds were discussed in  ref.~\cite{Cheung:2012nb} using a two-loop analysis for models with additional spin $0$ and $1/2$ electroweak singlet, doublet and triplets.  Vacuum stability in a scalar extended SM has been explored in refs.~\cite{Lebedev:2012zw, EliasMiro:2012ay} where the scalar is allowed to mix with the Higgs.  In a more elaborate model consisting of a complex scalar and two fermions, where one of the fermions serve as a DM particle, issues pertaining to vacuum stability have been discussed in ref.~\cite{Antipin:2013bya}. Vacuum metastability constraints in the context of seesaw models have been discussed in refs.~\cite{Chen:2012faa, Khan:2012zw}. Planck scale dynamics is expected to affect stability of EW vacuum. This has been demonstrated in refs.~\cite{Grzadkowski:2001vb, Branchina:2013jra, Lalak:2014qua, Branchina:2014usa} incorporating higher dimensional  operators  suppressed by $\mpl$ in the Higgs potential. Similar study in scalar DM model was performed in ref.~\cite{Eichhorn:2014qka}.

While the existing literature mainly concentrates on constraints ensuing from the stability of the Higgs effective potential, 
the possibility that the EW vacuum remains in the metastable state even after adding the scalar has not been discussed in depth. Although, in the context of SM, detailed studies of metastability exist in the literature~\cite{Bezrukov:2012sa, Degrassi:2012ry, Masina:2012tz, Buttazzo:2013uya}. 
Along the same line, in this paper, we extend studies of the metastable vacuum in the SM+$S$ model. In particular, we provide quantitative measures of metastability in terms of tunneling probability and confidence level plots.

The paper is organised as follows. In Section~\ref{chapt2} we deal with theoretical aspects of SM+$S$ model with a discussion on RGE running of various couplings and the Higgs effective potential. Tunneling probability of EW vacuum is discussed in Section~\ref{chapt3}, followed by presentation of phase diagrams in different parameter spaces in Section~\ref{chapt4}. Finally we conclude in Section~\ref{chapt5}.

\section{Effective potential and RGE running}
\label{chapt2}
An extra real scalar singlet field $S$, odd under $\Z_2$ symmetry, is added to the SM, providing a suitable candidate for dark matter. The corresponding Lagrangian density is given by,
\beq
{\cal L}_{S}= \frac{1}{2} (\partial_{\mu} S) (\partial^{\mu} S) -V_0^S
\eeq
with,
\beq
V_0^S= \frac{1}{2} \overline{m}_S^2 S^2 +\frac{\kappa}{2} |\Phi|^2 S^2 +
 \frac{\lambda_S}{4!} S^4 \nn \, ,
\eeq
where, 
\beq
\Phi=\begin{pmatrix}
G^+ \\ (\phi+v+iG^0)/\sqrt{2}
\end{pmatrix} \nn \, .
\eeq
After spontaneous EW symmetry breaking, DM mass $M_S$ is expressed as $M_S^2=\overline{m}_S^2  +\kappa v^2/2$.

SM tree level Higgs potential
\beq
V_0^{\rm SM}(\phi)=-\frac{1}{2} m^2 \phi^2 + \frac{1}{4} \lambda \phi^4
\label{v0}\\
\eeq
is augmented by the 
SM+$S$ one-loop Higgs potential in Landau gauge using $\MS$ scheme, which is written as
\beq
V_1^{{\rm SM}+S}(\phi)= V_1^{\rm SM}(\phi) + V_1^{S}(\phi)
\eeq
with~\cite{Lerner:2009xg, Gonderinger:2012rd},
\bea
V_1^{\rm SM}(\phi)&=&\sum_{i=1}^5 \frac{n_i}{64 \pi^2} M_i^4(\phi) \left[ \ln\frac{M_i^2(\phi)}{\mu^2(t)}-c_i\right] \nn \\
V_1^{S}(\phi)&=&\frac{1}{64 \pi^2} M_S^4(\phi) \left[ \ln\left(\frac{M_S^2(\phi)}{\mu^2(t)} \right)-\frac{3}{2} \right] \nn
\eea
where $n_i$ is the number of degrees of freedom. For scalars and gauge bosons, $n_i$ comes with a positive sign, whereas for fermions it is associated with a negative sign. Here $c_{H,G,f}=3/2$, $c_{W,Z}=5/6$ and $\mu(t)=M_Z \exp(t)$. $M_i(\phi)$ and $M_S(\phi)$ are given by
\beq
M_i^2(\phi)= \kappa_i(t) \phi^2(t)-\kappa_i^{\prime}(t)  \qquad {\rm and} \qquad M_S^2(\phi)= \overline{m}_S^2(t)  +\kappa(t) \phi^2(t)/2\, .
\nn \eeq
$n_i$, $\kappa_i$ and $\kappa_i^{\prime}$ can be found in eqn.~(4) in ref.~\cite{Casas:1994qy} (see also  refs.~\cite{Altarelli:1994rb, Casas:1994us, Casas:1996aq, Quiros:1997vk}). In this paper, in the Higgs effective potential, SM contributions are taken at two-loop level~\cite{Ford:1992pn, Martin:2001vx, Degrassi:2012ry, Buttazzo:2013uya}, whereas the scalar contributions are considered at one-loop only. 

For $\phi\gg v$, the effective potential can be approximated as
\beq
V_{\rm eff}^{{\rm SM}+S}(\phi) \simeq \lambda_{\rm eff}(\phi) \frac{\phi^4}{4}\, ,
\label{efflam}\eeq
with
\beq
\lambda_{\rm eff}(\phi) = \lambda_{\rm eff}^{\rm SM}(\phi) +\lambda_{\rm eff}^{S}(\phi)\, ,
\nn\eeq
where~\cite{Buttazzo:2013uya, Gonderinger:2012rd},
\bea
\lambda_{\rm eff}^{\rm SM}(\phi) &=& e^{4\Gamma(\phi)}
\left[ \lambda(\mu=\phi) + \lambda_{\rm eff}^{(1)}(\mu=\phi) +  \lambda_{\rm eff}^{(2)}(\mu=\phi)\right] \nn\\
 \lambda_{\rm eff}^{S}(\phi)&=&e^{4\Gamma(\phi)} \left[\frac{\kappa^2}{64 \pi^2}  \left(\ln\left(\frac{\kappa}{2}\right)-\frac{3}{2}\right ) \right]\, . \nn
\eea
Here 
\beq
\Gamma(\phi)=\int_{M_t}^{\phi} \gamma(\mu)\,d\,\ln\mu \, .\nn
\eeq
Anomalous dimension $\gamma(\mu)$ of the Higgs field takes care of  its wave function renormalisation.
As quartic scalar interactions do not contribute to wave function renormalisation at one-loop level, $S$ does not alter  $\gamma(\mu)$ of SM and anomalous dimension of $S$ is zero~\cite{Gonderinger:2009jp}. 
The expressions for the one- and two-loop quantum corrections $\lambda_{\rm eff}^{(1,2)}$ in SM can be found in ref.~\cite{Buttazzo:2013uya}.  
All running coupling constants are evaluated at $\mu=\phi$.

For RGE running, we use three-loop SM beta functions~\cite{Mihaila:2012fm,  Chetyrkin:2012rz, Zoller:2012cv, Chetyrkin:2013wya, Zoller:2013mra, Buttazzo:2013uya} for $g_1$, $g_2$, $g_3$, $y_t$ and $\lambda$. For $\mu\geq M_S$, $\beta_\lambda$ receives a correction~\cite{Haba:2013lga} $\kappa^2/2$ at one-loop due to $S$. 
The beta functions for the new physics parameters $\kappa$ and $\lambda_S$ are given by~\cite{Haba:2013lga, Clark:2009dc, Davoudiasl:2004be}
 \begin{align}
  &\beta_\kappa
    =\left\{
      \begin{array}{ll}
       0 & \mbox{for }\mu<M_S\\
       \kappa\left[4\kappa+12\lambda+6y^2-\frac{3}{2}(g'{}^2+3g^2)+\lambda_S\right] & \mbox{for }\mu\geq M_S
      \end{array}
     \right. \, ,   \\
  &\beta_{\lambda_S}
    =\left\{
      \begin{array}{ll}
       0 & \mbox{for }\mu<M_S\\
       3\lambda_S^2+12\kappa^2 & \mbox{for }\mu\geq M_S
      \end{array}
     \right. \, . 
 \end{align}
 $\overline{m}_S$ also evolves with energy. But as the beta functions of other parameters do not involve $\overline{m}_S$, we do not consider its beta function in this discussion.  Here, new physics effects are included in the RGEs at one-loop only.

\subsection{RGE running from $\mu=M_t$ to $\mpl$}

We evaluate the coupling constants at the highest mass scale of the SM namely the top quark mass $M_t$ and then run them according to the  RGEs up to the Planck scale. To know their values at $M_t$, one needs to take into account various threshold corrections up to  $M_t$~\cite{Sirlin:1985ux, Melnikov:2000qh, Holthausen:2011aa}. All coupling constants are expressed in terms of pole masses~\cite{Bezrukov:2012sa}.  To calculate $g_1 (M_t)$ and $g_2 (M_t)$, one-loop RGEs are enough. For $g_3 (M_t)$,  we first use three-loop RGE running of $\alpha_s$ with five flavours excluding the top quark, and, then the effect of top is included using prescriptions of an effective field theory. The leading term in four-loop RGE for $\alpha_S$ is also taken into account.  Amongst all Yukawa couplings, the running of $y_t$ is the most significant. $\MS$ $y_t$ is related to the top pole mass $M_t$ by the matching condition $y_t(M_t) = \frac{\sqrt{2} M_t}{v} \left( 1+\delta_t (M_t) \right)$. In $\delta_t (M_t)$, we take into account three-loop QCD, one-loop electroweak, and two-loop ${\cal O}(\alpha \alpha_S)$ corrections. Similarly, the relation between $\MS$ $\lambda$ and Higgs pole mass $M_H$ is given by $\lambda(M_t) = \frac{M_H^2}{2 v^2} \left(1+\delta_H (M_t) \right)$. In $\delta_H (M_t)$, we consider one-loop electroweak, two-loop ${\cal O}(\alpha \alpha_S)$, two-loop ${\cal O}(y_t^4 g_3 ^2)$ and two-loop ${\cal O}(y_t^6)$  corrections. 
The loop effects considered in these matching conditions are comparable to refs.~\cite{Bezrukov:2012sa, Degrassi:2012ry}. After knowing the values of various coupling constants at $M_t$, we use full three-loop SM RGEs and one-loop RGEs for the scalar $S$ to run them up to $\mpl$.

Using these matching conditions, in Table~\ref{table1} we find different coupling constants at $M_t = 173.1$~GeV and at $\mpl$ with $M_H = 125.66$~GeV and $\alpha_s(M_Z) = 0.1184$. As we are considering running all the way up to the Planck scale, the values at very high energies are extremely sensitive to the initial values at $M_t$. Hence, significant places of decimals are provided to enable the reader to reproduce the running. 
\begin{table}[h!]
\begin{center}
    \begin{tabular}{ | c | c |  c | c | c | c |}
    \hline
     $\mu$ (GeV) & $g_1$ & $g_2$ & $g_3$ & $y_t$ & $\lambda$\\
    \hline
     $M_t$ & 0.358725 & 0.648184 & $1.16449$ & $0.935644$ & $0.126971$  \\
     \hline
     $\mpl$ & 0.476006 & 0.506548 & $0.488986$ & $0.384124$ & $-0.0123196$  \\
     \hline
             \end{tabular}
    \caption{\textit{Values of all {\rm SM} coupling constants at $M_t$ and $\mpl$ for $M_t=173.1$~GeV, $M_H=125.66$~GeV and $\alpha_S(M_Z)=0.1184$. }}
    \label{table1}
\end{center}
\end{table}

For SM+$S$, the running depends on the extra parameters $M_S$, $\kappa(M_Z)$ and $\lambda_S(M_Z)$. Assuming the values of SM parameters at $M_t$ as given in Table~\ref{table1},  we present values of all parameters at $\mpl$ in SM+$S$ model in Table~\ref{table2}, for two different sets of $M_S$, $\kappa(M_Z)$ and $\lambda_S(M_Z)$. The first set stands for our benchmark point, as described later. For this choice, $\lambda(\mpl)$ is negative. The second set is chosen such that  $\lambda(\mpl)=0$.  Note that as we do not consider  running of new physics parameters till $M_S$,  in our case it is all the same to specify these parameters either at $M_Z$ or at $M_t$. 
\begin{table}[h!]
\begin{center}
    \begin{tabular}{ | c | c | c | c |  c | c | c | c | c | c |}
    \hline
    $M_S$ & $\kappa(M_Z)$ & $\lambda_S(M_Z)$ & $g_1$ & $g_2$ & $g_3$ & $y_t$ & $\lambda$ & $\kappa$ & $\lambda_S$\\
\hline
    620 & 0.185 & 1 & 0.478 & 0.506 & 0.487   & 0.382 & $-0.0029$ & 0.424 &  4.66 \\
    \hline
    795 & 0.239 & 0.389 & 0.478 & 0.506 & 0.487 & 0.382 & 0 & 0.412 & 1  \\         \hline
    \end{tabular}
    \caption{\textit{ Values of all {\rm SM+}$S$ coupling constants at $\mpl  = 1.2 \times 10^{19}$~GeV with $M_t=173.1$~GeV, $M_H=125.66$~GeV and $\alpha_S(M_Z)=0.1184$.}}
    \label{table2}
\end{center}
\end{table}

 \begin{figure}[h!]
 \begin{center}
 \includegraphics[width=3.7in,height=3.1in, angle=0]{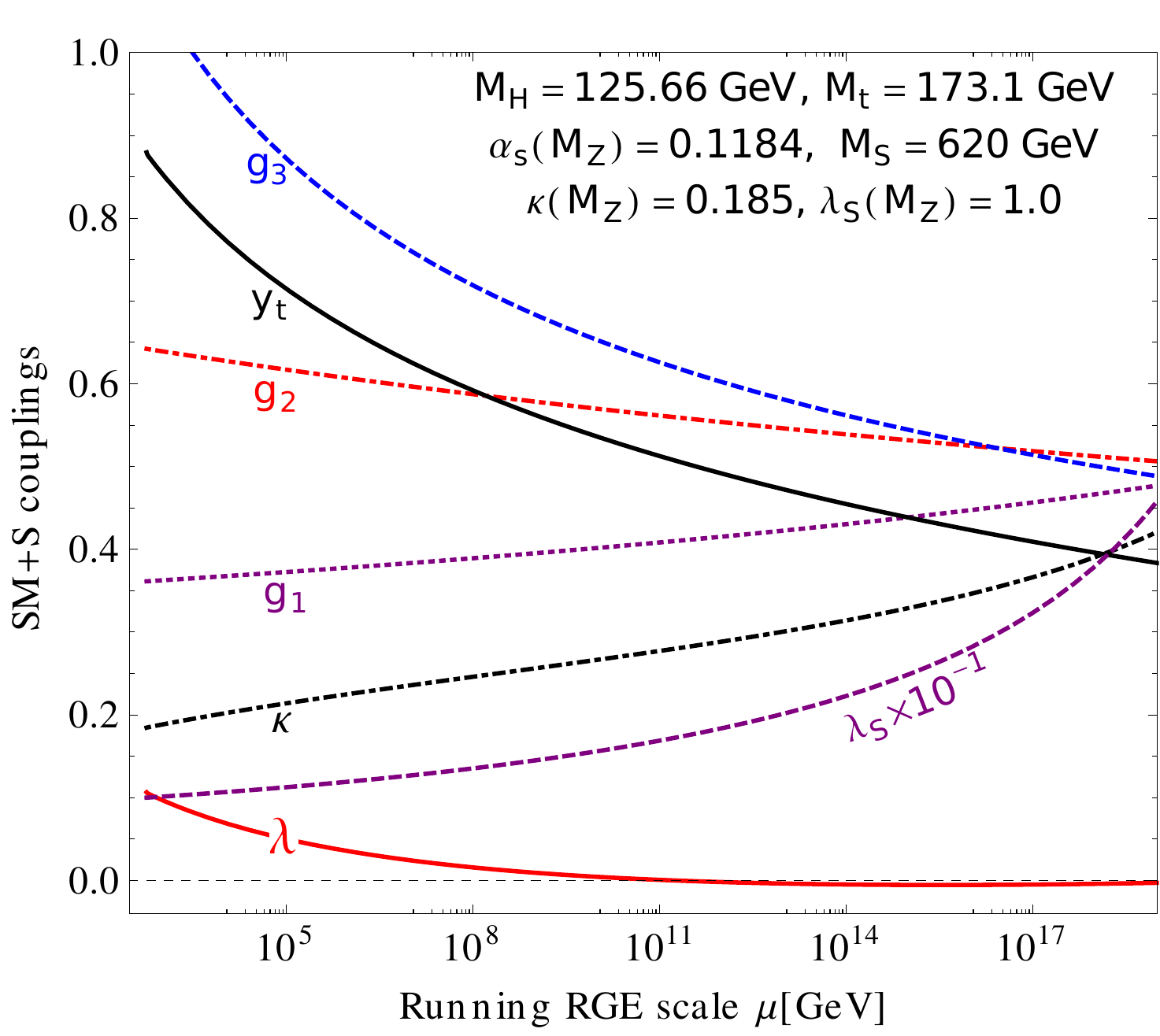}
 \caption{\label{fig:SMScalar} \textit{  {\rm SM+}$S$ RG evolution of the gauge couplings $g_1,g_2,g_3$, top Yukawa coupling $y_t$, Higgs self-quartic coupling $\lambda$, Higgs portal coupling $\kappa$ and scalar self-quartic coupling $\lambda_S$ in $\MS$ scheme. } }
 \end{center}
 \end{figure}
In the SM, the gauge couplings $g_1,g_2,g_3$, top Yukawa coupling $y_t$ do not vanish at $\mpl$. The same is true in SM+$S$ model~\cite{Gabrielli:2013hma}, since running of these couplings are hardly affected by $S$, as displayed in Fig.~\ref{fig:SMScalar}.  Higgs portal coupling $\kappa$ and scalar self-quartic coupling $\lambda_S$ increase with energy. The rise of $\lambda_S$ is so rapid that it may render the theory nonperturbative at higher energies. For example, if $\lambda_S(M_Z)>1.3$, $\lambda_S$ becomes nonperturbative before $\mpl$.  Higgs self-quartic coupling $\lambda$ 
 also gets affected by inclusion of $S$. But  the change is not visible in Fig.~\ref{fig:SMScalar}. 
 \begin{figure}[h]
 \begin{center}
 \subfigure[]{
 \includegraphics[width=2.8in,height=2.8in, angle=0]{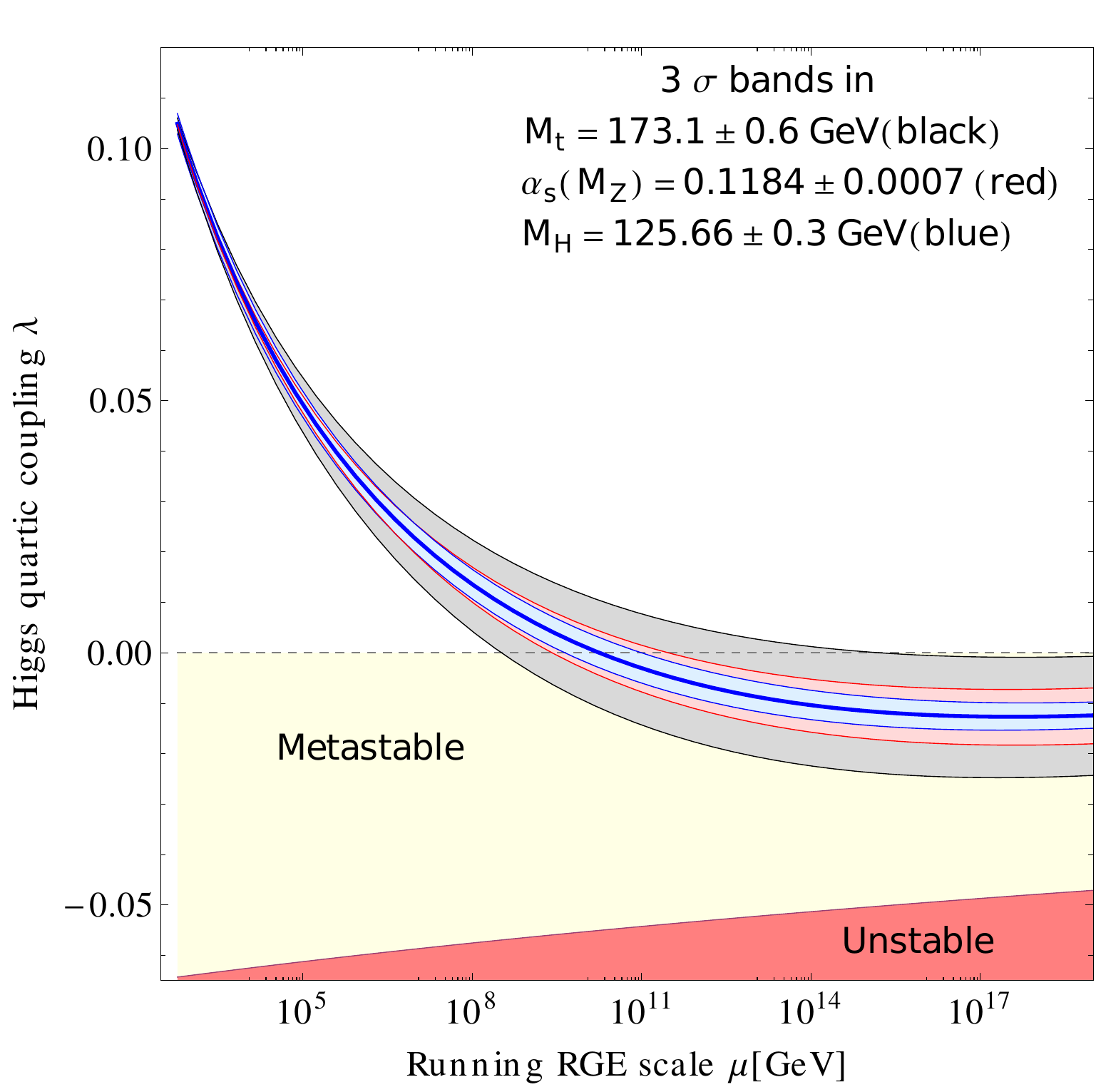}}
 \subfigure[]{
 \includegraphics[width=2.8in,height=2.8in, angle=0]{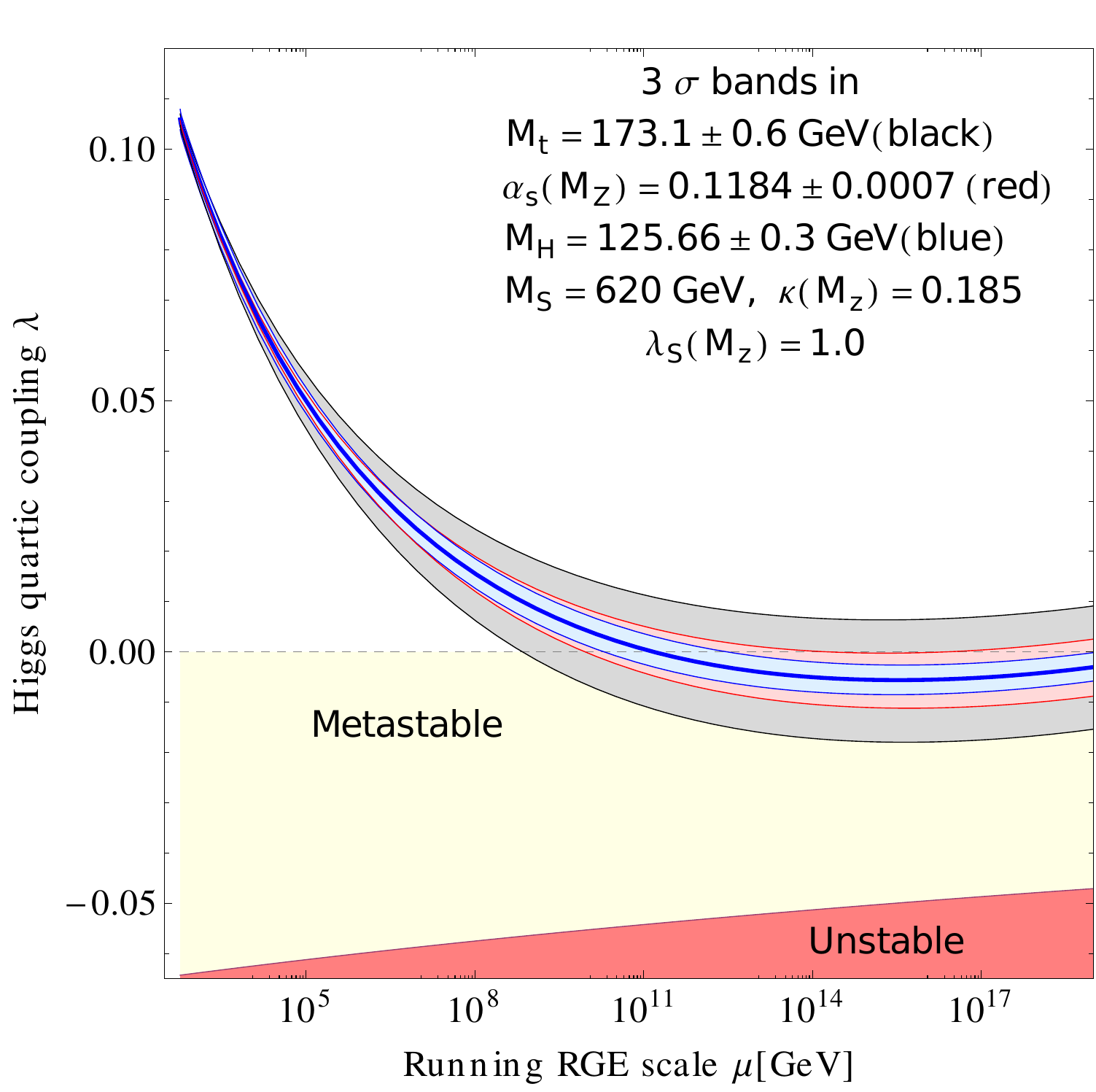}}
 \caption{\label{fig:lambda} \textit{$({\bf a})$ RG evolution of $\lambda$ in the {\rm SM}. $({\bf b})$ RG evolution of $\lambda$ in the {\rm SM+}$S$ for our benchmark point. In both panes $3\sigma$ bands for $M_t,M_H$ and $\alpha_s(M_Z)$ are displayed.} }
 \end{center}
 \end{figure}

As the issue of stability hinges on the value of $\lambda$ at higher energies, we focus on the running of $\lambda$ both in SM and SM+$S$ models in Fig.~\ref{fig:lambda}. In SM, $\lambda$ becomes vanishingly small and negative at energies before $\mpl$, signifying the possible presence of new physics. $\lambda$ crosses zero at a scale $\Lambda_I$, where the instability starts setting in. 
Considering central values for $M_t$, $M_H$ and $\alpha_S(M_Z)$,  $\Lambda_I\sim 1.9 \times 10^{10}$~GeV in the SM. We display the same curve for SM+$S$ in Fig.~\ref{fig:lambda}(b), for a {\em benchmark point}  $M_S = 620$ GeV, $\kappa(M_Z)=0.185$ and $\lambda_S (M_Z)=1$. We observe that the behaviour of $\lambda$ running might change significantly, modifying $\Lambda_I$ to $1.68 \times 10^{11}$~GeV. It has the potential to push out the EW vacuum from metastability to a stable vacuum. The benchmark point is chosen keeping in mind that the new physics effects are clearly visible, yet the vacuum is still in metastable state. This point also satisfies the Planck and WMAP imposed DM relic density constraint $\Omega_{\rm DM}h^2=0.1198\pm 0.0026$~\cite{Ade:2013zuv}. We use {\tt FeynRules}~\cite{Alloul:2013bka} along with {\tt micrOMEGAs}~\cite{Belanger:2010gh, Belanger:2013oya} to compute relic density of scalar DM in SM+$S$ model.  
 \begin{figure}[h!]
 \begin{center}
 \subfigure[]{
 \includegraphics[width=2.8in,height=2.8in, angle=0]{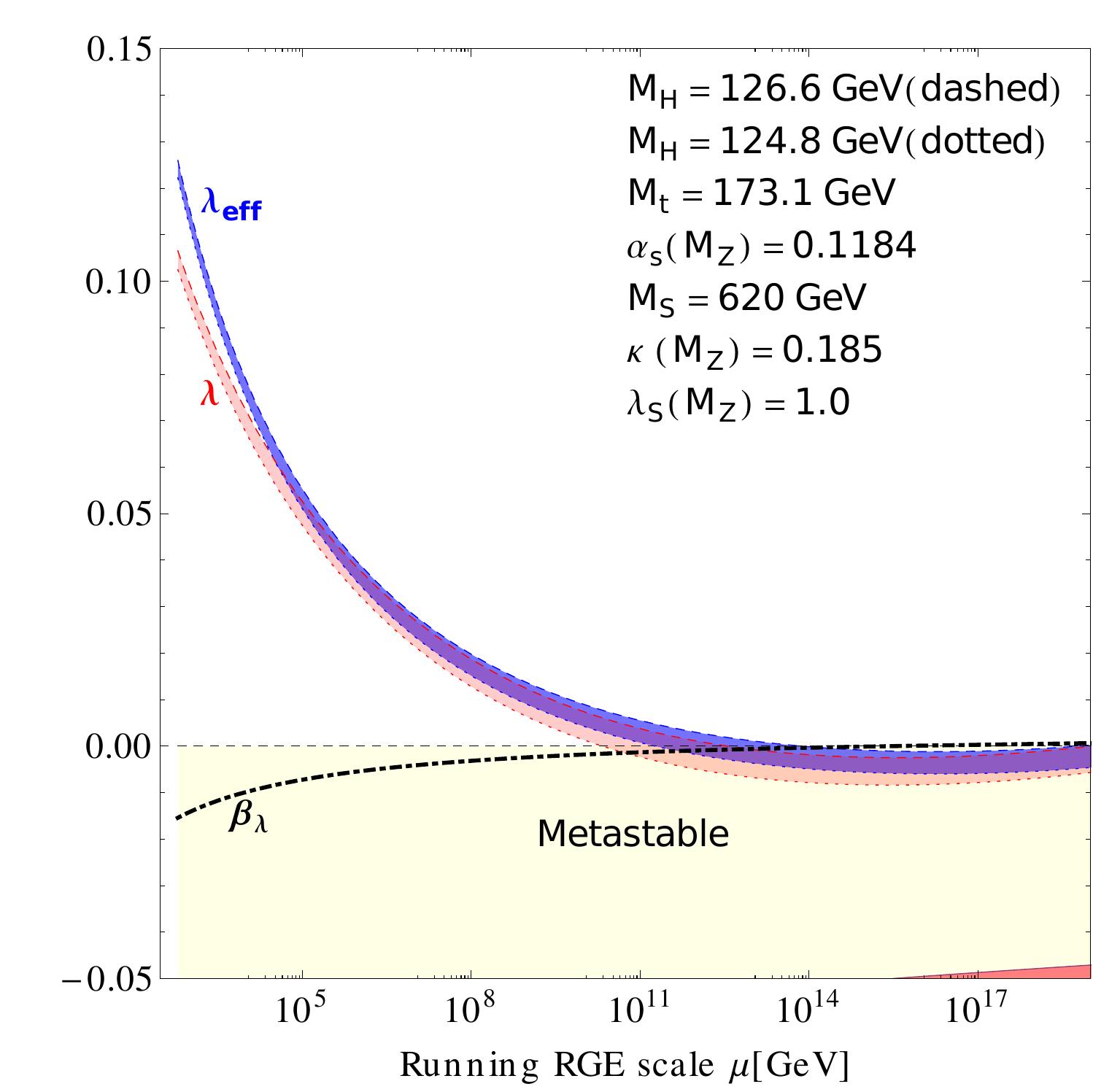}}
 \subfigure[]{
 \includegraphics[width=2.8in,height=2.8in, angle=0]{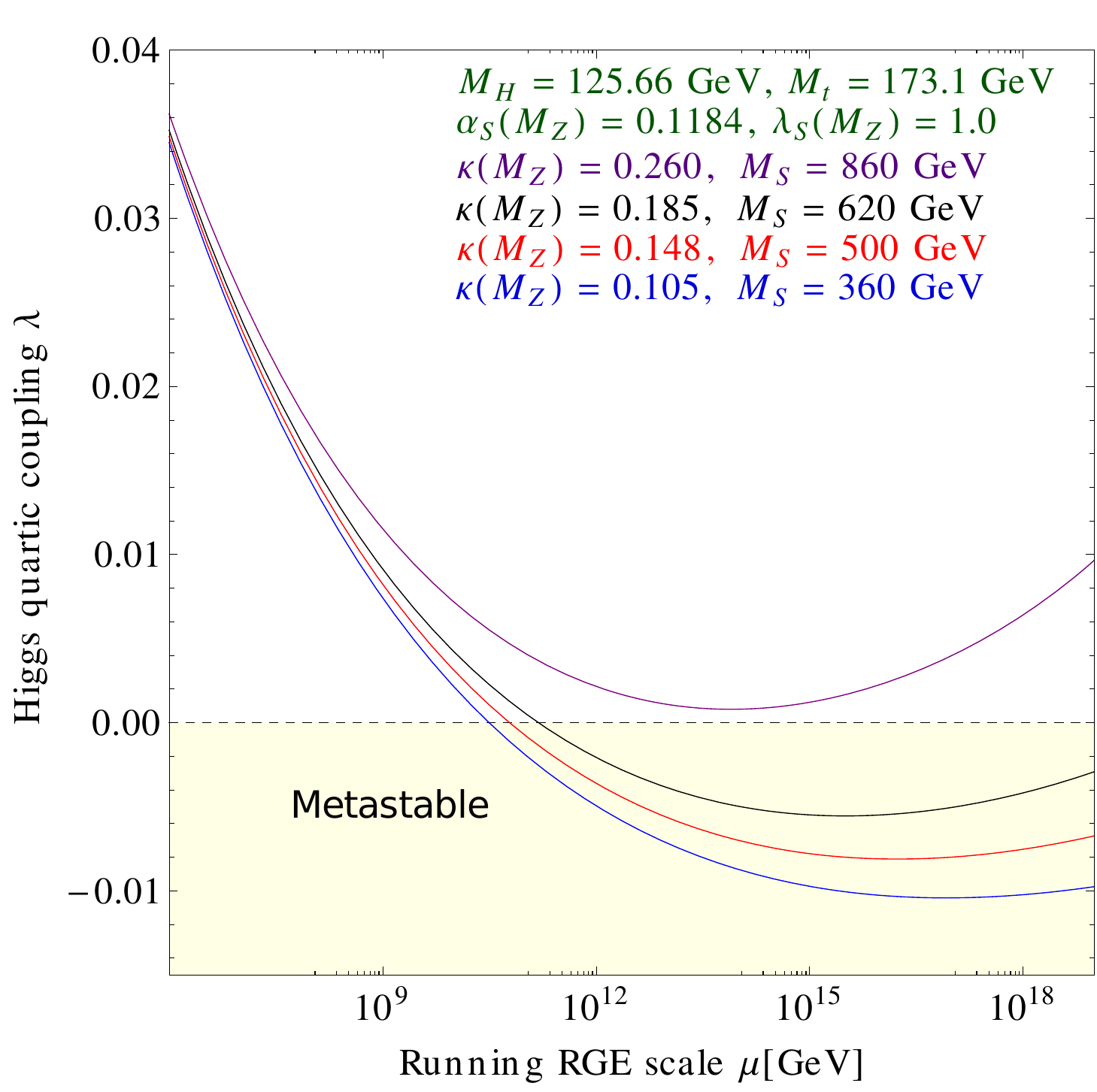}}
 \caption{\label{fig:lambdaef} \textit{$({\bf a})$ RG evolution of $\lambda$ (red band),  $\lambda _{\rm eff}$ (blue band)  and  $\beta_\lambda$ (dot-dashed black) in {\rm SM+}$S$ for our benchmark point. $({\bf b})$~Evolution of  $\lambda$ for different $\kappa(M_Z)$. Each $\kappa(M_Z)$ corresponds to a specific $M_S$ to satisfy DM relic density $\Omega_{\rm DM} h^2 \approx$ 0.1198.} }
 \end{center}
 \end{figure}

As defined in eqn.~(\ref{efflam}), $\lambda_{\rm eff}$ differs from $\lambda$ as it takes care of loop corrections. In the SM, in ref.~\cite{Degrassi:2012ry} it was shown that the difference $\lambda_{\rm eff}-\lambda$ is always positive and negligible near $\mpl$. We see in Fig.~\ref{fig:lambdaef}(a) that {\rm SM+}$S$ exhibit similar features. However, the instability scale $\Lambda_I$ changes significantly if we choose to work with $\lambda_{\rm eff}$ instead of $\lambda$: In SM, the instability scale changes to $1.25 \times 10^{11}$~GeV and in {SM+}$S$, it becomes $1.7\times 10^{12}$~GeV. We also plot $\beta_\lambda$ to show that at high energies, $\lambda$, $\lambda_{\rm eff}$ and  $\beta_\lambda$ all seem to vanish. 

In Fig.~\ref{fig:lambdaef}(b) we display RGE running of $\lambda$ for various new physics parameters to explicitly demonstrate that as $\kappa(M_Z)$ increases, for a given energy, $\lambda$ assumes a higher value~\cite{Kadastik:2011aa, Chen:2012faa}. Finally, for some parameter space, $\lambda$ never turns negative, implying stability of the EW vacuum. It happens due to the $\kappa^2/2$ term in $\beta_\lambda$.  Due to this positive contribution, the presence of the scalar never drives EW vacuum towards instability.  Next, we will calculate the tunneling probability to demonstrate stability issues with EW vacuum.

\section{Tunneling probability}
\label{chapt3}

The present data on $M_H$ and $M_t$ indicate that the Universe might be residing in a false vacuum, waiting for a quantum tunneling to a true vacuum lying close to the Planck scale.

The vacuum decay probability of EW vacuum at the present epoch is given by~\cite{Coleman:1977py, Isidori:2001bm, Buttazzo:2013uya}
\beq
{\cal P}_0=0.15 \frac{\Lambda_B^4}{H_0^4} e^{-S(\Lambda_B)}
\label{prob}\\
\eeq
where,
\beq
S(\Lambda_B)=\frac{8\pi^2}{3|\lambda(\Lambda_B)|}
\label{action}\\
 \eeq
 is the action of bounce of size $R=\Lambda_B^{-1}$. Tunneling is dominated by that $R$ for which  $S(\Lambda_B)$  is minimum, \ie when $\lambda(\Lambda_B)$ is minimum, implying $\beta_\lambda(\Lambda_B)=0$. For a given set of model parameters, in our analysis, $\Lambda_B$ is determined from the constraining conditions like $\beta_\lambda(\Lambda_B)=0$, $\lambda(\Lambda_B)=0$ \etc whenever applicable.

 \begin{figure}[h!]
 \begin{center}
\subfigure[]{
 \includegraphics[width=2.8in,height=2.8in, angle=0]{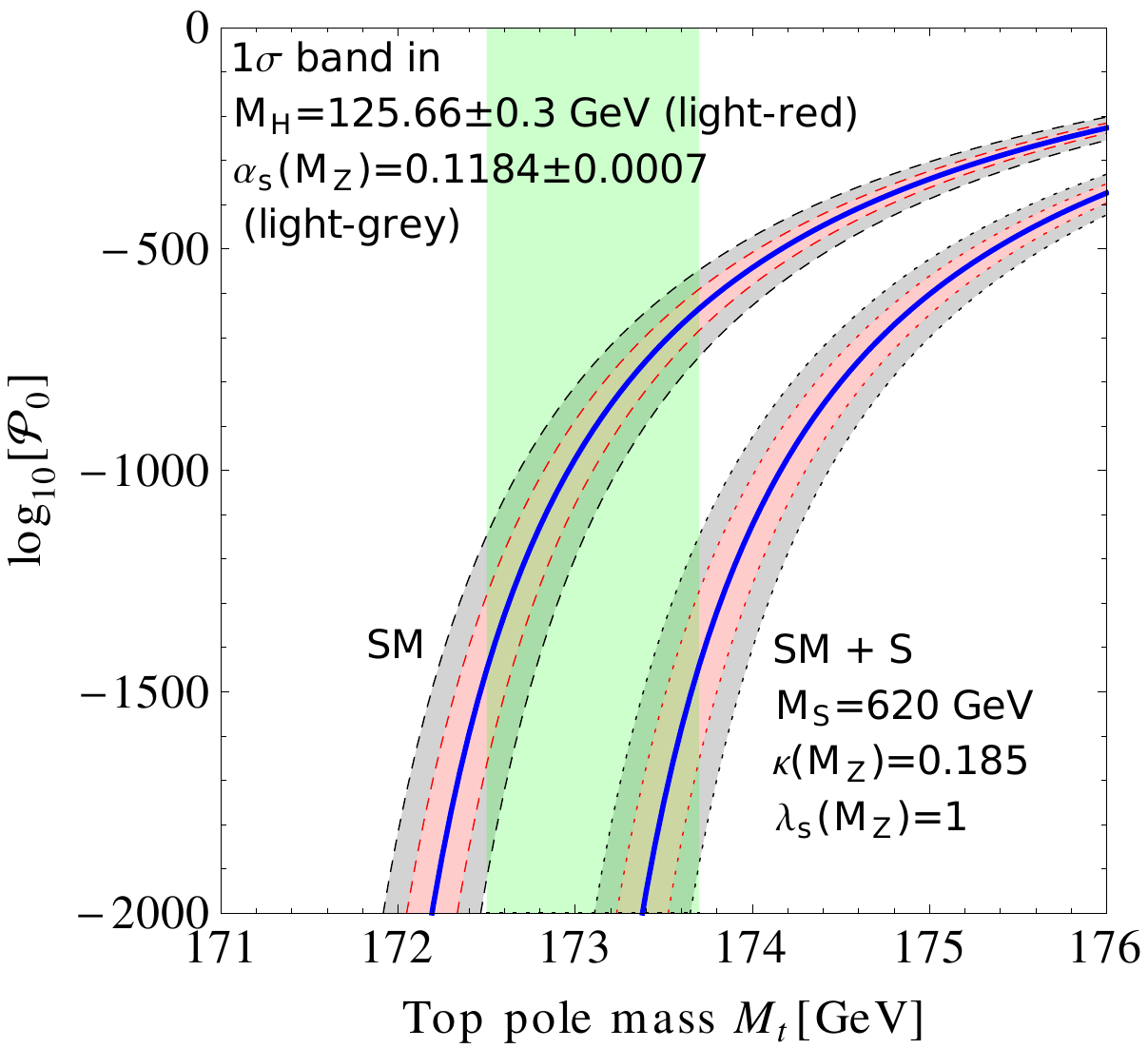}}
 \subfigure[]{
 \includegraphics[width=2.8in,height=2.8in, angle=0]{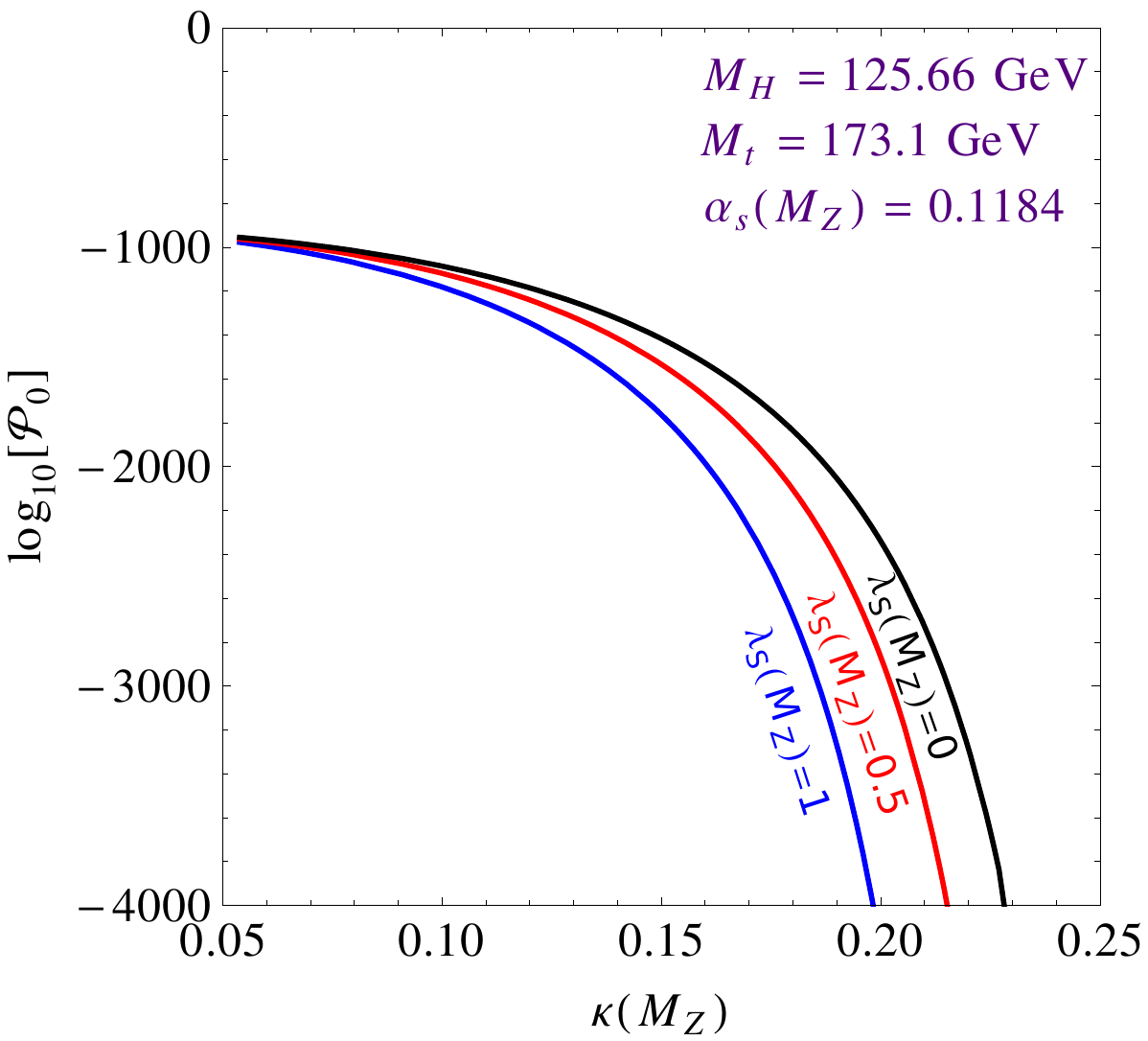}}
 \caption{\label{fig:Tun} \textit{ $({\bf a})$ Tunneling probability ${\cal P}_0$ as a function of  $M_t$. The left band corresponds to {\rm SM} and the right one to {\rm SM+}$S$ for our benchmark point. Light-green band stands for $M_t$ at $\pm 1\sigma$. $({\bf b})$ ${\cal P}_0$ as a function of $\kappa(M_Z)$ for various $\lambda_S(M_Z)$.} }
 \end{center}
 \end{figure}
The action $S$ receives one-loop correction $\Delta S$, which is rather insignificant~\cite{Isidori:2001bm} when we set the running scale $\mu = 1/R$. The correction due to gravitation $\Delta S_G = {256 \pi^3 \Lambda_B^2 }/{45 \left(\mpl \lambda \right)^2 }$~\cite{Coleman:1980aw, Isidori:2007vm} is relevant only when $\Lambda_B \sim \mpl$. In this work, we neglect both of these contributions.

${\cal P}_0 < 1$ corresponds to a vacuum decay lifetime greater than the lifetime of the Universe, $\tau_U \simeq 0.96/H_0 \sim 13.7$ billion years, which in turn implies~\cite{Isidori:2001bm},
\beq
\lambda(\Lambda_B) > \lambda_{\rm min}(\Lambda_B)=\frac{-0.06488}{1-0.00986 \ln\left( {v}/{\Lambda_B} \right)}\,.
\label{lammin}
\eeq
In passing, we note the following:
\begin{itemize}
\item
If $\lambda(\Lambda_B)>\frac{4\pi}{3}$, $|\kappa|>8\pi$, $|\lambda_S| > 8\pi$\footnote{Note that these numbers differ from the ones used in ref.~\cite{Haba:2013lga}.} then the theory is nonperturbative~\cite{Lee:1977eg, Cynolter:2004cq, Modak:2013jya}. 
\item
If $\lambda(\Lambda_B)>0$, then the vacuum is stable. 
\item
If $0>\lambda(\Lambda_B)>\lambda_{\rm min}(\Lambda_B)$, then the vacuum is metastable. 
\item
If $\lambda(\Lambda_B)<\lambda_{\rm min}(\Lambda_B)$, then the vacuum is unstable. 
\item
If $\lambda_S<0$, the potential is unbounded from below along the $S$-direction. 
\item
If $\kappa<0$, the potential is unbounded from below along a direction in between $S$ and $H$.  
\end{itemize}
In Fig.~\ref{fig:Tun} we have plotted tunneling probability ${\cal P}_0$ as a function of  $M_t$. To calculate ${\cal P}_0$, we first find the minimum value of $\lambda_{\rm eff}$ and put the same in eqn.~(\ref{action}) to minimise $S(\Lambda_B)$.  The right band corresponds to the tunneling probability for our benchmark point. 
For comparison, we plot the same for SM as the left band in Fig.~\ref{fig:Tun}(a). 1$\sigma$ error bands in $\alpha_S$ and in $M_H$ are also displayed. The error due to $\alpha_S$ is clearly more significant\footnote{There is a typo in Fig.~7 of ref.~\cite{Buttazzo:2013uya}. The bands due to $\alpha_S$ and $M_H$ have been interchanged.} than that due to $M_H$. 
We observe that for a given $M_t$, these new physics effects lower the tunneling probability. It bolsters our earlier observation that scalar $S$ helps the EW vacuum to come out of metastability.  We demonstrate this in Fig.~\ref{fig:Tun}(b), where we plot 
${\cal P}_0$ as a function of  $\kappa(M_Z)$ for different choices of $\lambda_S(M_Z)$, assuming central values for $M_H$, $M_t$ and $\alpha_S$. We see that for low values of $\kappa(M_Z)$, ${\cal P}_0$ tends to coincide with its SM value. For a given $\kappa(M_Z)$, for higher $\lambda_S(M_Z)$, ${\cal P}_0$ gets smaller, making the EW vacuum more stable. 

\section{Phase diagrams}
\label{chapt4}
The stability of EW vacuum is best displayed with the aid of phase diagrams. We present phase diagrams in different parameter planes for our model. 
\subsection{$M_H - M_t$ phase diagram}
To draw the phase diagram in $M_H-M_t$ plane, we need to identify regions pertaining to EW vacuum stability, metastability and unstable regions. The line separating the stable region from the metastable one is obtained when the two vacua are at the same depth, implying $\lambda(\Lambda_B)=\beta_\lambda(\Lambda_B)=0$. The unstable region is differentiated from the metastability region by the boundary line where $\beta_\lambda(\Lambda_B)=0$ along with $\lambda(\Lambda_B)=\lambda_{\rm min}(\Lambda_B)$, defined in eqn.~(\ref{lammin}).
 \begin{figure}[h]
 \begin{center}
 \subfigure[]{
 \includegraphics[width=2.8in,height=2.8in, angle=0]{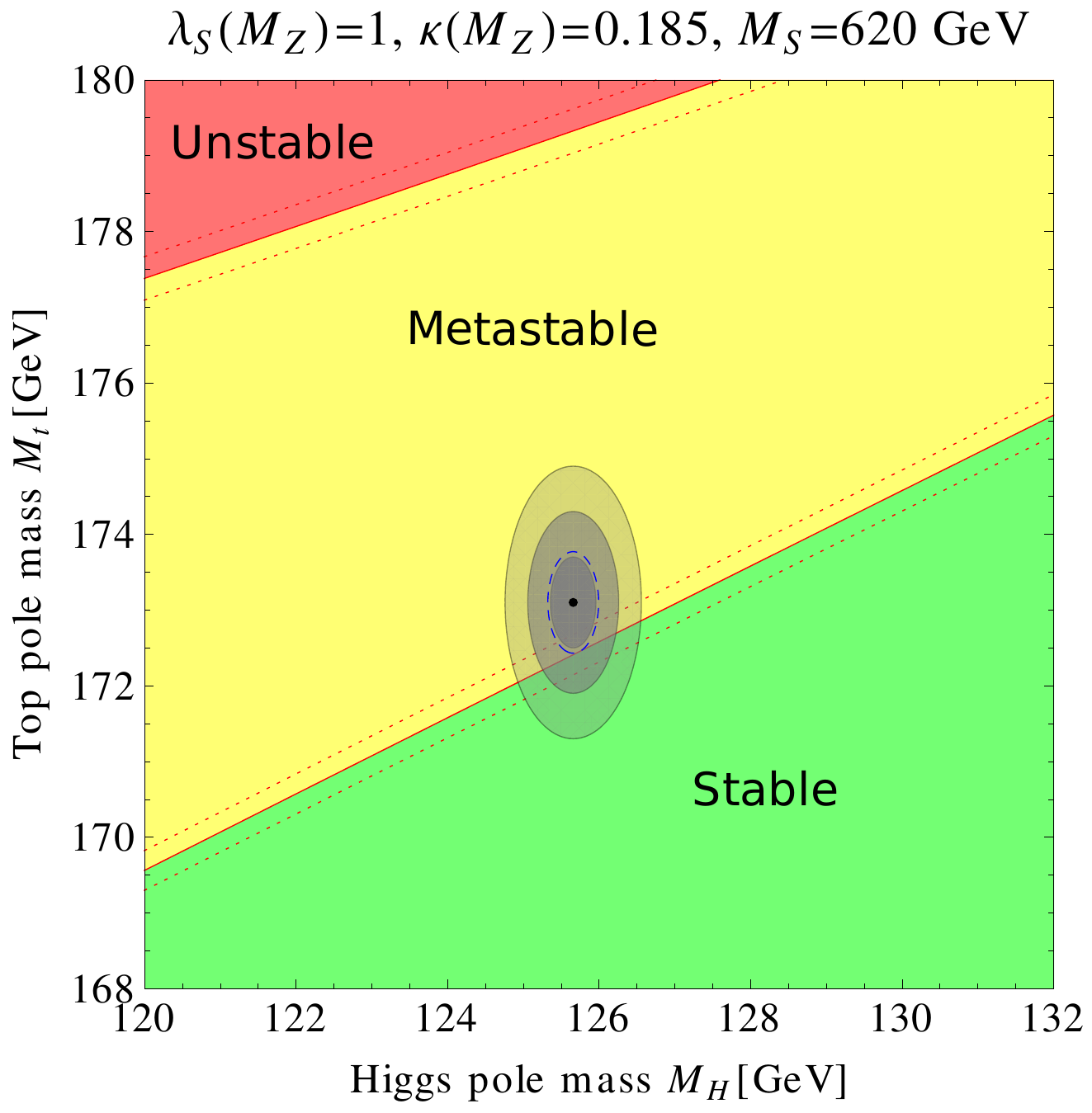}}
 \subfigure[]{
 \includegraphics[width=2.8in,height=2.8in, angle=0]{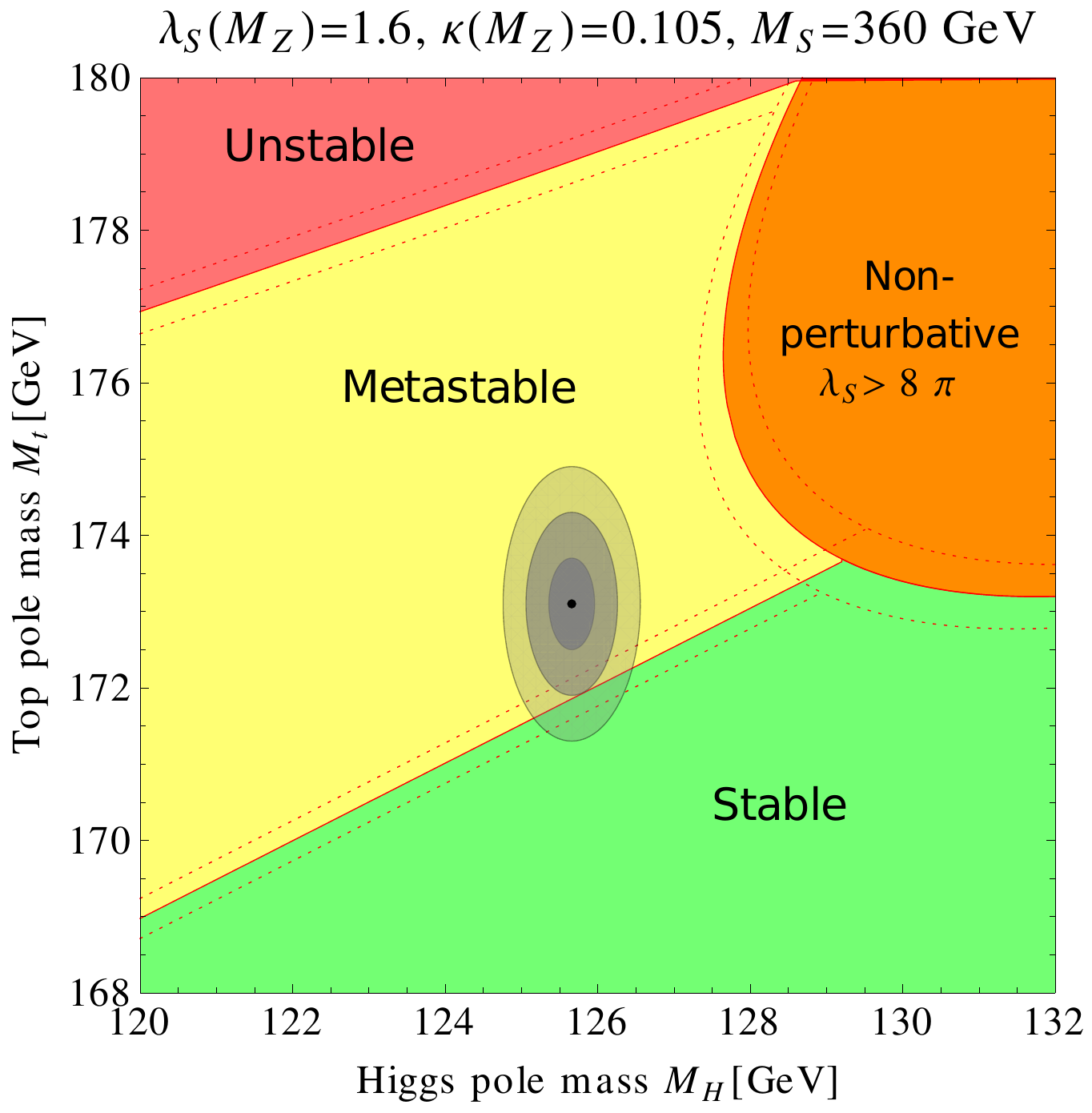}}
 \caption{\label{fig:Mt_Mh} \textit{$({\bf a})$ In {\rm SM+}$S$, regions of absolute stability~(green), metastability~(yellow), instability~(red) of the EW vacuum in the $M_H - M_t$ plane phase diagram is presented for our benchmark point $M_S = 620$~GeV, $\kappa(M_Z)=0.185$ and $\lambda_S (M_Z)=1$. 
 $({\bf b})$ Similar plot for $M_S = 360$~GeV, $\kappa(M_Z)= 0.105$ and $\lambda_S (M_Z) = 1.6$. The orange region  corresponds to nonperturbative zone for $\lambda_S$.
The three boundary lines (dotted, solid and dotted red) correspond to $\alpha_s(M_Z)=0.1184 \pm 0.0007$. The grey areas denote the experimentally favoured zones for $M_H$ and $M_t$ at $1$, $2$ and  $3\sigma$. } }
 \end{center}
 \end{figure}

In the SM, the phase diagram in $M_H-M_t$ plane is given in  refs.~\cite{Degrassi:2012ry, Buttazzo:2013uya}. Our results agree with them. Given the measured errors on $M_t=173.1\pm 0.6$~GeV and $M_H$, the SM phase diagram indicates that the stability of EW vacuum is excluded at $\sim 3\sigma$\footnote{In a mass dependent renormalisation scheme, such exclusion happens at $3.5\sigma$~\cite{Spencer-Smith:2014woa}.}.  However, the extra scalar in our model modifies these findings. 

We present the phase diagram in the $M_H - M_t$ plane for SM+$S$ in Fig.~\ref{fig:Mt_Mh}. For our benchmark point, we see that the boundaries shift towards higher values of $M_t$, so that the EW vacuum stability is excluded only at 1.1$\sigma$, indicated by the blue-dashed ellipse. For  $M_S = 360$~GeV, $\kappa(M_Z)= 0.105$ and $\lambda_S (M_Z) = 1.6$, the plot is redrawn to highlight the fact that $\lambda_S$ might turn out to be too large, so that the theory becomes nonperturbative (marked as the orange region in Fig.~\ref{fig:Mt_Mh}(b)). Here EW vacuum stability is excluded at 2$\sigma$. All these boundaries separating various stability regions in the phase diagram depend on $\alpha_S$. 1$\sigma$ bands for the same is also displayed in these figures.

\subsection{Confidence level of vacuum stability}

 \begin{figure}[h]
 \begin{center}
 {
 \includegraphics[width=3in,height=2.8in, angle=0]{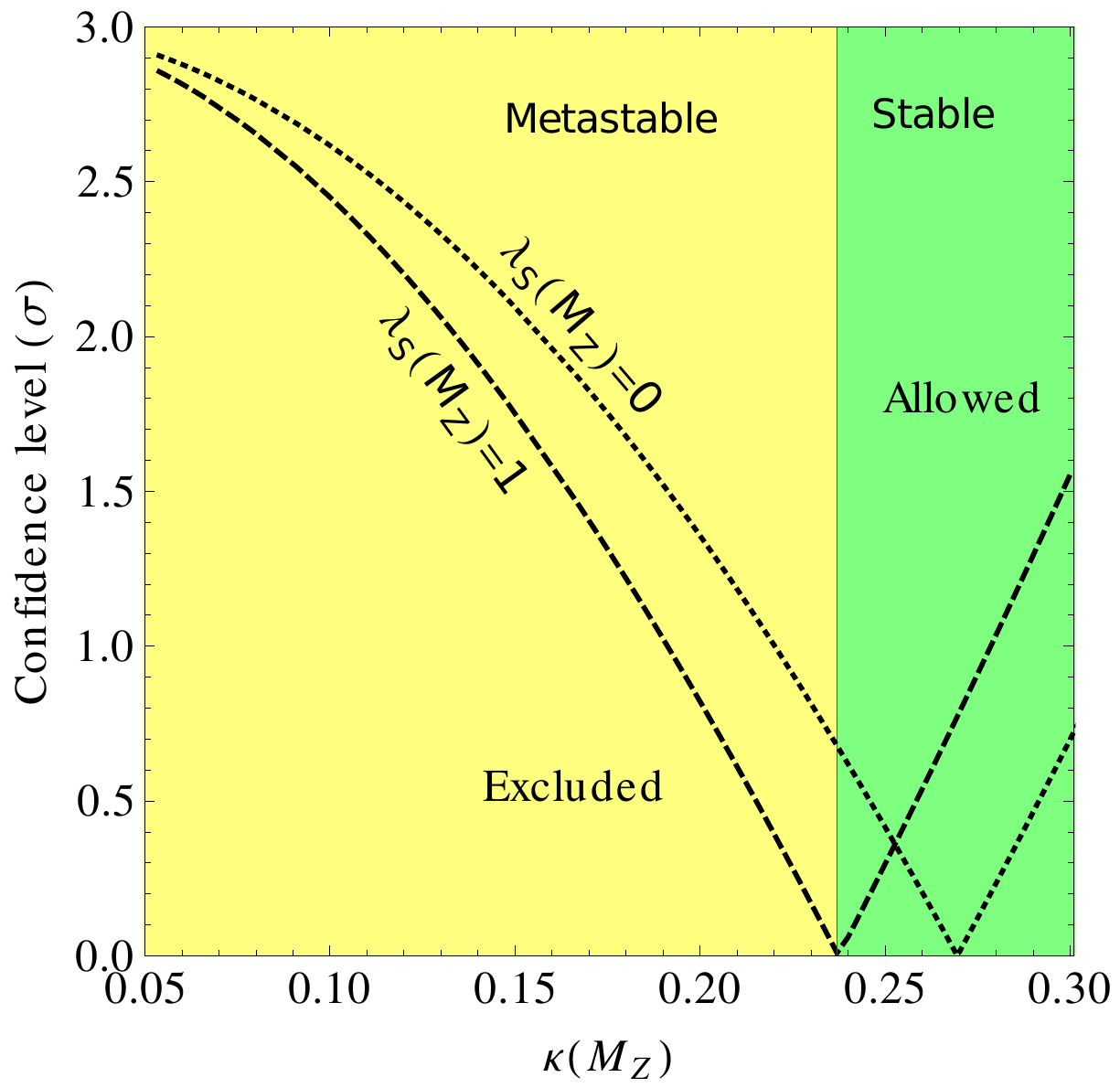}}
 \caption{\label{fig:kappaconfidence} \textit{Dependence of confidence level (one-sided) at which EW vacuum stability is excluded/allowed on $\kappa(M_Z)$. Regions of absolute stability (green) and metastability (yellow) of EW vacuum are shown.} }
 \end{center}
 \end{figure}
As new physics effects do change the stability of EW vacuum, it is important to show the change in the confidence level at which stability is excluded or allowed. In Fig.~\ref{fig:kappaconfidence}, we plot confidence level against $\kappa(M_Z)$ for $M_t=173.1$~GeV, $M_H=125.66$~GeV and $\alpha_S(M_Z)=0.1184$. $M_S$ is dictated by $\kappa(M_Z)$ to satisfy $\Omega_{\rm DM} h^2\approx 0.1198$. For $\lambda_S(M_Z)=1$, we see that the EW vacuum becomes stable for  $\kappa(M_Z)=0.24$ onward. For a lower $\lambda_S(M_Z)$, this point shifts to a higher value. If $\lambda_S(M_Z)=0$, stability is assured for $\kappa(M_Z)\ge 0.27$. Note that as $\kappa$ dependence in RGE running of $\lambda$ creeps in through the term $\kappa^2/2$ in $\beta_\lambda$, the stability strongly depends on $\kappa(M_Z)$. However, as  $\beta_\lambda$ depends on $\lambda_S$ only {\it via} $\kappa$ running, although $\lambda_S$ running is relatively strong, the stability of EW vacuum does not change appreciably when $\lambda_S(M_Z)$ is varied from $0$ to $1$.

\subsection{$M_t-\alpha_S(M_Z)$ phase diagram}

 \begin{figure}[h]
 \begin{center}
 {
 \includegraphics[width=3in,height=2.8in, angle=0]{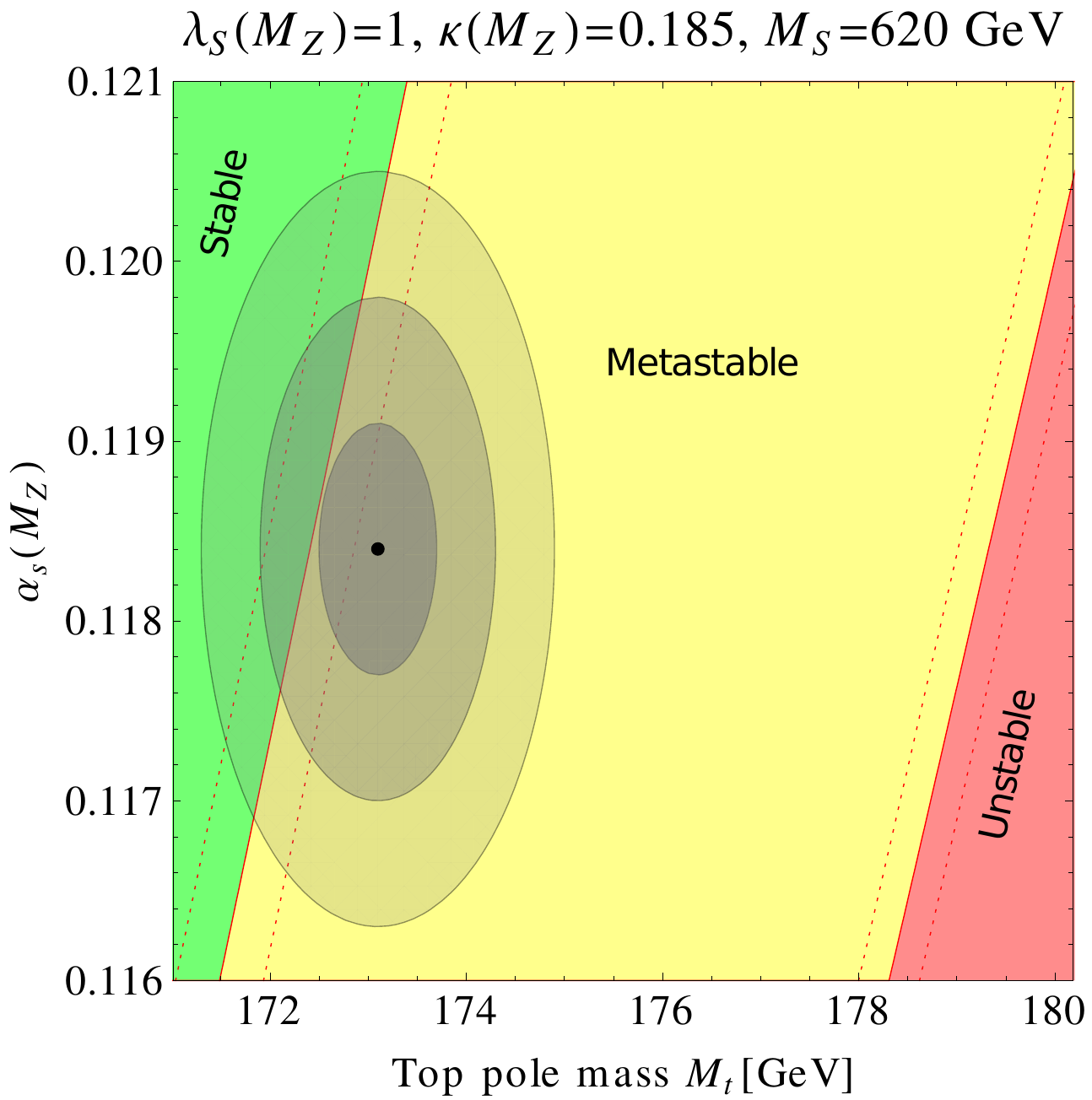}}
 \caption{\label{fig:Alpha_Mt} \textit{Phase diagram in $M_t-\alpha_S(M_Z)$ plane in {\rm SM+}$S$ model for our benchmark point. Regions of absolute stability~(green), metastability~(yellow), instability~(red) of the EW vacuum are marked. The dotted lines correspond to $\pm 3\sigma$ variation in $M_H$ and the grey areas denote the experimental allowed region for $M_t$ and $\alpha_S(M_Z)$ at $1$, $2$ and  $3\sigma$.} }
 \end{center}
 \end{figure}

Given the sizable error on $\alpha_S(M_Z)$, it is instructive to draw the phase diagram in the $M_t-\alpha_S(M_Z)$ plane as well. This diagram for SM is available in refs.~\cite{Bezrukov:2012sa, EliasMiro:2011aa}. We present the same in Fig.~\ref{fig:Alpha_Mt} for our model using our benchmark point.  With increase of $\kappa(M_Z)$ and/or $\lambda_S(M_Z)$,  the boundaries between different stability regions shift towards right, allowing the EW vacuum to be more stable.

\subsection{Asymptotic safety}
 \begin{figure}[h]
 \begin{center}
 \subfigure[]{
 \includegraphics[width=2.8in,height=2.8in, angle=0]{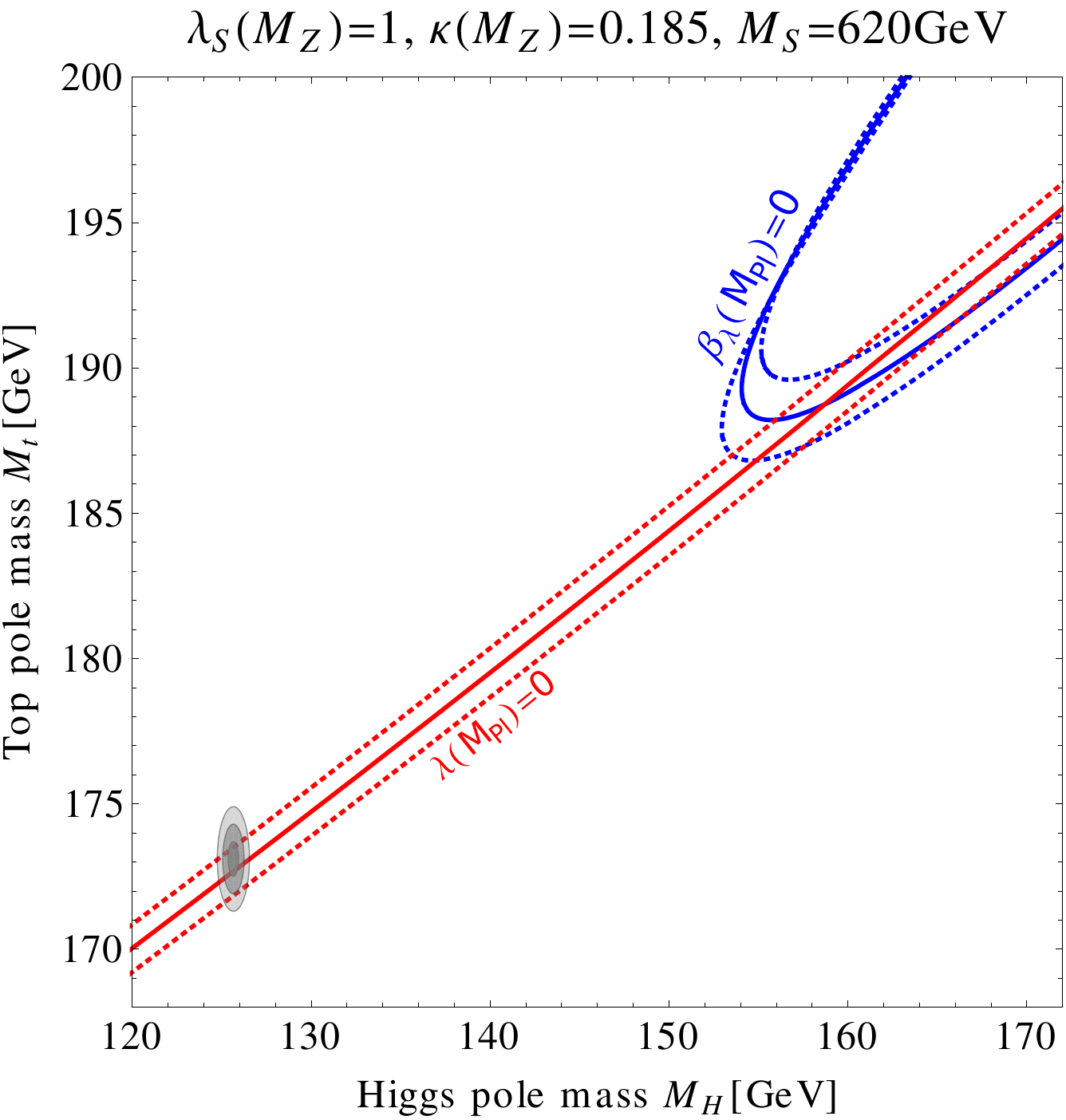}}
 \subfigure[]{
 \includegraphics[width=2.8in,height=2.8in, angle=0]{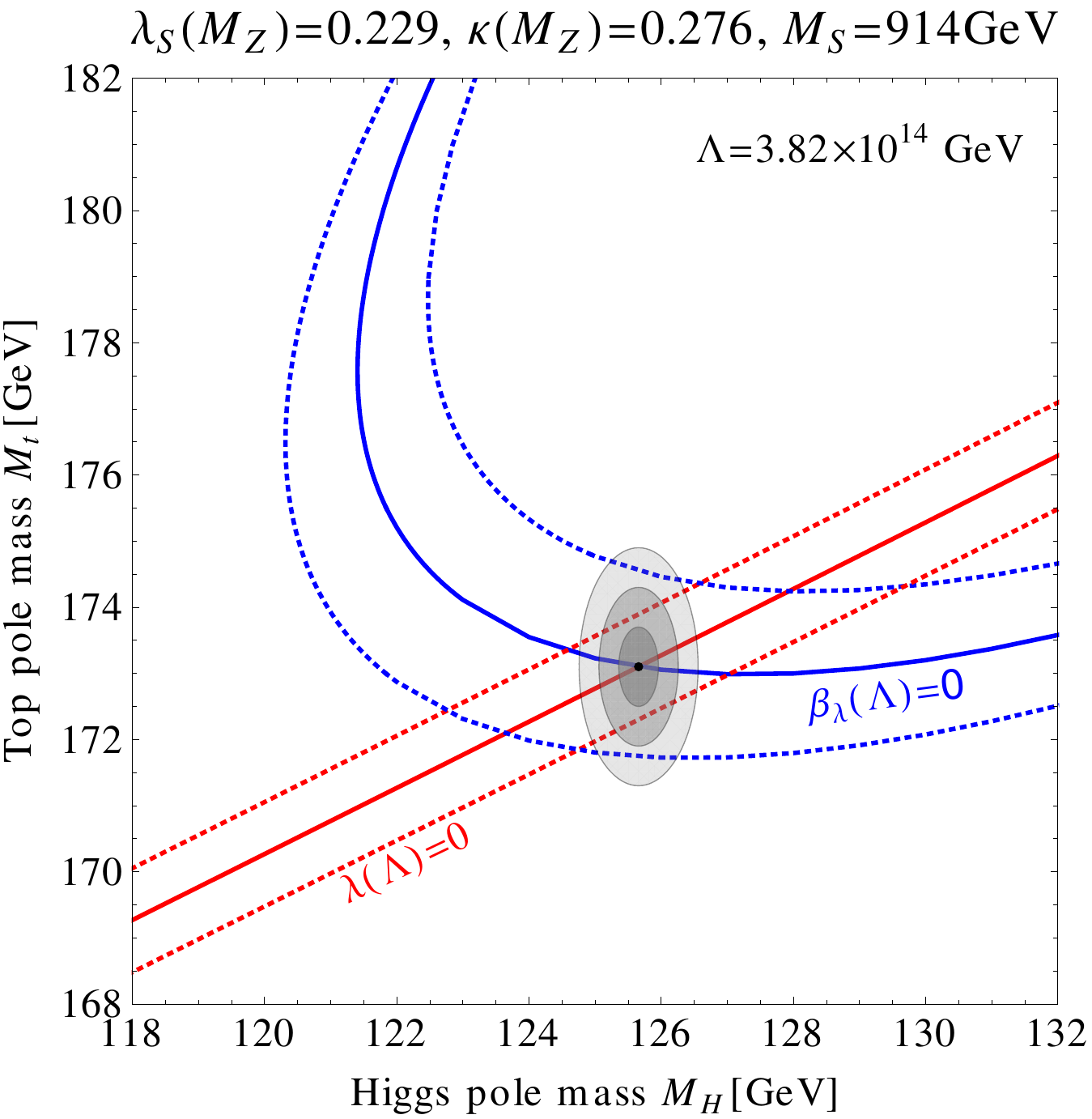}}
 \caption{\label{fig:asymp} \textit{$({\bf a})$ Contour plot for $\lambda(\mpl) =0$ (red line) and  $\beta_{\lambda}(\mpl)=0$ (blue line) for our benchmark point. $({\bf b})$ Similar plot for $\lambda ({\Lambda}) =0$ and  $\beta_{\lambda}(\Lambda)=0$, where $\Lambda=3.8\times10^{14}$~GeV. Dotted lines correspond to $\pm 3\sigma$ variation in $\alpha_s(M_Z)$. The grey areas denote the experimental allowed region for $M_H$ and $M_t$ at $1$, $2$ and  $3\sigma$.} }
 \end{center}
 \end{figure}

Shaposhnikov and Wetterich predicted~\cite{Shaposhnikov:2009pv} mass of the Higgs boson of $126$~GeV imposing the constraint $\lambda(\mpl)=\beta_{\lambda}(\mpl)=0$, in a scenario known as asymptotic safety of gravity. As mentioned before, this corresponds to two degenerate vacua. In ref.~\cite{Degrassi:2012ry}, it was clearly shown that the present error in $M_t$ and $M_H$ does not allow this condition to be realised in SM. In SM+$S$ model, the situation worsens~\cite{Haba:2013lga} and we  demonstrate this in Fig.~\ref{fig:asymp}(a) for our benchmark point. In presence of the scalar, the values of $M_t$ and $M_H$, required to satisfy this condition, are pushed far away from the experimentally favoured numbers: For our benchmark point this condition is satisfied at  $M_H=140.8$ GeV and  $M_t=179.5$ GeV.

However, it is possible to meet this condition at a lower energy than at $\mpl$. In Fig.~\ref{fig:asymp}(b),  it is demonstrated that at a different point in the parameter space: $M_S = 914$~GeV, $\kappa(M_Z)=0.276$ and $\lambda_S (M_Z)=0.229$,  the condition $\lambda ({\Lambda}) =\beta_{\lambda}(\Lambda)=0$ is indeed satisfied at $\Lambda=3.8\times10^{14}$ GeV and is also consistent with experimentally allowed range for $M_t$ and $M_H$. The value of $\Lambda$ decreases with  $\lambda_S (M_Z)$ and $\kappa(M_Z)$. The corresponding value of $M_S$ is chosen to satisfy relic density of DM constraints. 
Also, it is difficult to simultaneously satisfy Veltman condition at $\mpl$~\cite{Haba:2013lga}. All these observations indicate that some new physics could be operational at very high energies  to take care of such issues.

\subsection{$\kappa(M_Z)-M_S$ phase diagram}
 \begin{figure}[h]
 \begin{center}
 \includegraphics[width=3in,height=2.8in, angle=0]{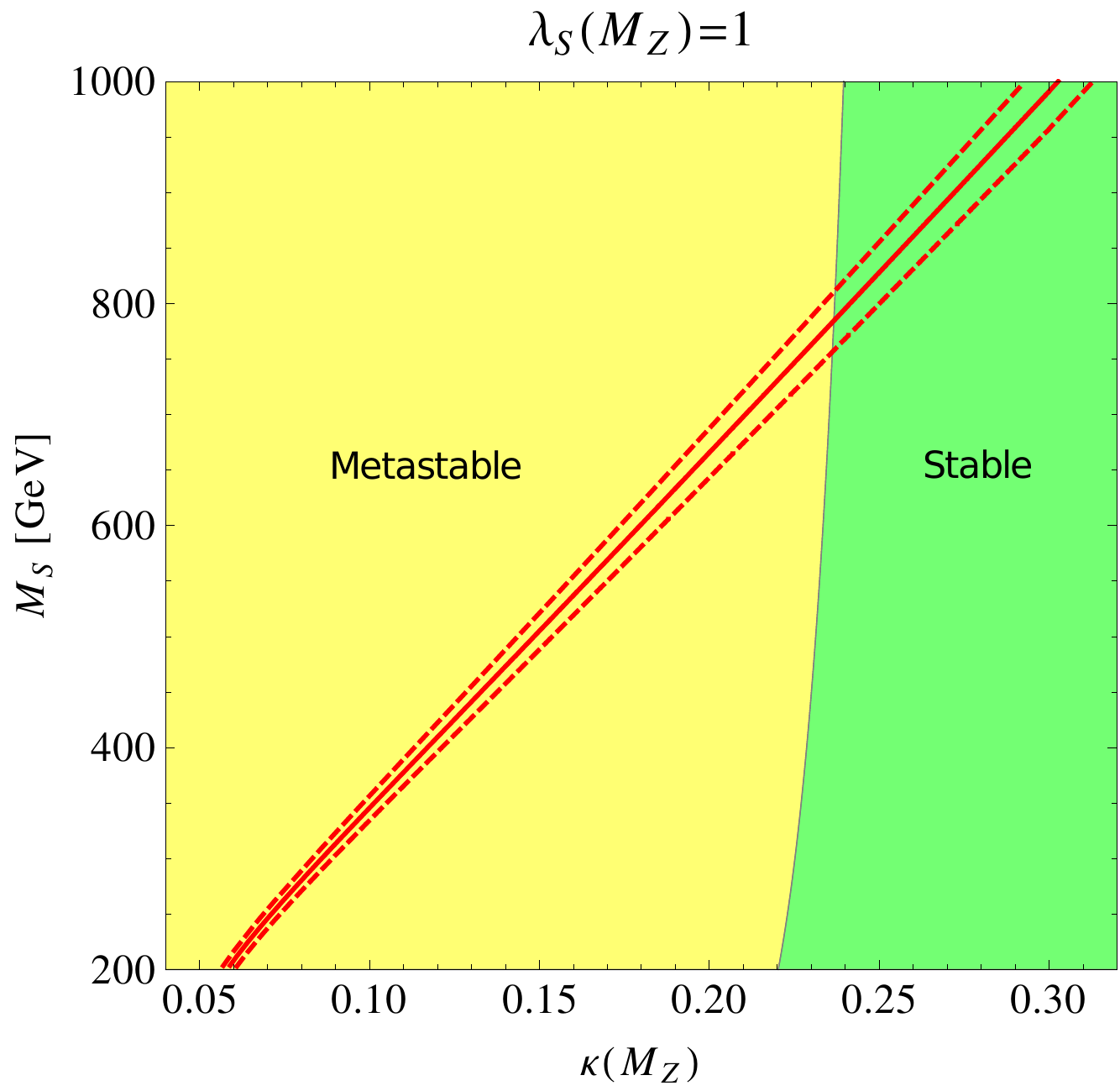}
 \caption{\label{fig:Mskappa} \textit{ Phase diagram in $\kappa(M_Z)-M_S$ plane. Regions of absolute stability (green), metastability (yellow) of the EW vacuum for $\lambda_S (M_Z) = 1$ are shown. The red solid line corresponds to $\Omega_{\rm DM} h^2 \approx$ 0.1198. The red dashed lines correspond to 3$\sigma$ error on $\Omega_{\rm DM} h^2$.} }
 \end{center}
 \end{figure}
The phase diagram for $\kappa(M_Z)-M_S$ plane is displayed in Fig.~\ref{fig:Mskappa}. As addition of the scalar does not drive the EW vacuum towards instability, there is no unstable region marked on the plot. Between the dashed lines, we mark the allowed region ensuing from relic density constraints. For $M_S> 100$~GeV, $SS\ra W^+ W^-$~\cite{McDonald:1993ex} dominates over other DM annihilation channels. Under the approximation $M_S\gg M_W, M_H$, in the non-relativistic limit,  
\beq
\sigma(SS\ra W^+ W^-) \propto \frac{\kappa^2}{M_S^2} \, .
\eeq 
Hence, to satisfy relic density constraints,  $\kappa(M_Z)$ depends linearly on $M_S$ as shown in Fig.~\ref{fig:Mskappa}. 

\subsection{$\kappa(M_Z)- M_H$ phase diagram}
 \begin{figure}[h!]
 \begin{center}
 \includegraphics[width=3.3in,height=2.8in, angle=0]{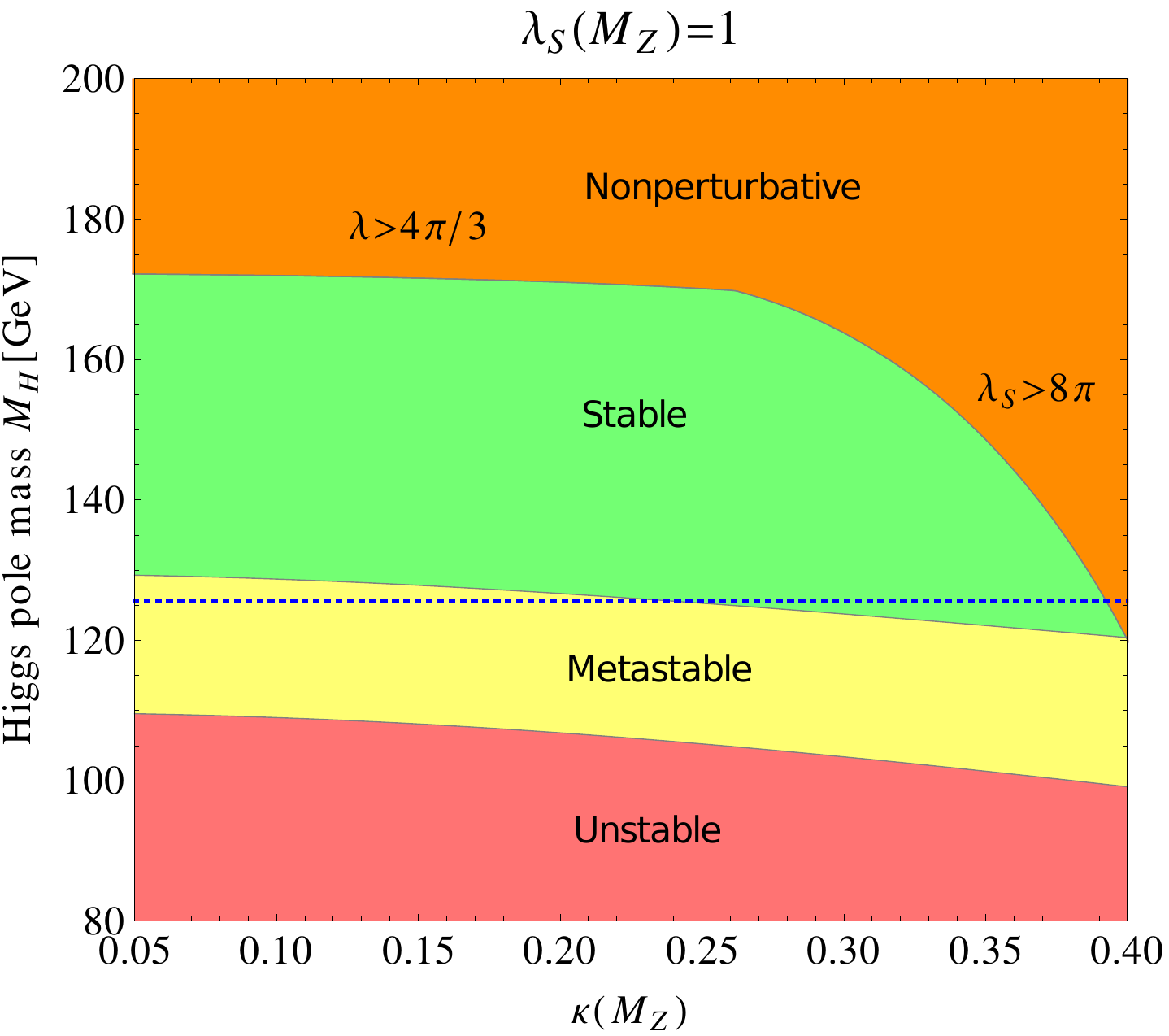}
 \caption{\label{fig:Mhkappa} \textit{ Phase diagram in $\kappa(M_Z)- M_H$ plane. Regions of absolute stability (green), metastability (yellow) and instability (red) of the EW vacuum for $\lambda_S (M_Z) = 1$ are displayed. $\lambda$ and/or $\lambda_S$ are/is nonperturbative in the orange region. The blue dashed line corresponds to $M_H=125.66$~GeV. } }
 \end{center}
 \end{figure}
 In SM, the vacuum stability bound 
 \beq
 M_H~{\rm [GeV]}>129.46+1.12\left( \frac{M_t~{\rm [GeV]} -173.1}{0.6} \right) 
 - 0.56 \left(\frac{\alpha_S(M_Z)-0.1184}{0.0007}\right)
  \eeq
  is obtained from the requirement $\lambda=\beta_{\lambda}=0$. The metastability bound
\beq
 M_H~{\rm [GeV]}>109.73+1.84\left( \frac{M_t~{\rm [GeV]} -173.1}{0.6} \right) 
 - 0.88 \left(\frac{\alpha_S(M_Z)-0.1184}{0.0007}\right)
  \eeq  
comes from the requirement $\beta_\lambda=0$ and $\lambda=\lambda_{\rm min}$. To ensure perturbativity, we demand $\lambda(\mpl)<\frac{4\pi}{3}$, which leads to
\beq
 M_H~{\rm [GeV]}<172.23+0.36\left( \frac{M_t~{\rm [GeV]} -173.1}{0.6} \right) 
 - 0.12 \left(\frac{\alpha_S(M_Z)-0.1184}{0.0007}\right)\, .
  \eeq  
Now, let us see what happens in SM+$S$ model. In this context, 
change in $M_H$ bounds with respect to $\kappa(M_Z)$ was considered in refs.~\cite{Gonderinger:2009jp, Eichhorn:2014qka} for different cut-off scales, considering stability aspects only. As shown in Fig.~\ref{fig:Mhkappa}, in presence of the scalar $S$, these bounds shift to lower values for larger $\kappa(M_Z)$. For large values of $\kappa(M_Z)$, depending on our choice of $\lambda_S$ at $M_Z$,  $\lambda_S(\mpl)$ may become so large that the theory becomes nonperturbative. This imposes further constraints on the parameter space, shown as the curved line representing $\lambda_S(\mpl)=8\pi$. As before, for a given $\kappa(M_Z)$, $M_S$ is chosen in such a way that  $\Omega_{\rm DM} h^2\approx 0.1198$ for $M_H=125.66$~GeV. However, in the plot, as $M_H$ changes,  $\Omega_{\rm DM} h^2$ also changes. But this variation is contained within 3$\sigma$.

\section{Discussion and conclusion}
\label{chapt5}
According to the standard model of particle physics, the electroweak vacuum is lying right in between stability and instability, as if, it is ready to tunnel into a regime of absolute instability. Nevertheless, the transition time required for this is safely beyond the present lifetime of the Universe. Still, the question is, what prompts such a near-criticality? In this paper, we did not try to find an answer to this question. But we explored the validity of this question in the context of an extended model containing a singlet scalar dark matter.  

Near-criticality is best explained with the help of phase diagrams. Refs.~\cite{Bezrukov:2012sa, Degrassi:2012ry, Buttazzo:2013uya} demonstrated quite a few of them to explain metastability in SM. We made a similar endeavour in the SM+$S$ model. We included NNLO corrections in our SM calculations, but the effects due to the scalar $S$ were incorporated at the level of one-loop only. 

We chose a dark matter model to illustrate changes in EW vacuum stability as apart from neutrino masses, the presence of dark matter in the Universe is the most striking signature of new physics beyond the standard model of particle physics. SM+$S$ is the simplest dark matter model one can work with. The present work deals with DM mass of $200$~GeV or more. The work can be extended to lower DM masses as well. However, in this case, to satisfy DM relic density constraints, $\kappa(M_Z)$ will also be small, leading to little deviation in the EW vacuum metastability from the SM case.  

Non-observation of any signature pertaining to supersymmetry or extra dimensions or any other TeV scale physics at LHC has led us to cast doubt on our take on the naturalness issues associated with the Higgs sector. 
It has led to a possibility that SM is not an effective field theory, but a fundamental theory of Nature, valid up to the Planck scale. To explain dark matter, here we assumed the singlet scalar extended SM reigns up to $\mpl$ as well.  One need not be concerned  about the hierarchy problem as we assume that Planck scale physics somehow takes care of that. In any case, as new scalars are of mass within a TeV, we are not introducing any new hierarchy problem in traditional sense. As the Higgs potential seems to vanish around $\mpl$, it has led to speculations~\cite{Holthausen:2013ota, Froggatt:1995rt, Shaposhnikov:2009pv, Hashimoto:2013hta, Foot:2013hna, Ibe:2013rpa, Kobakhidze:2014afa, Haba:2014sia} regarding the underlying dynamics operational at very high energies.

In SM+$S$, for some specific choice of parameter space, the scalar $S$ can rescue the EW vacuum from metastability, making it absolutely stable, so that $\lambda$ never turns negative. But as the DM direct detection experiments or collider searches are yet to confirm the exact nature of DM candidate, we consider the parameter space allowed by DM relic density constraints, dictated by the CMBR experiments such as Planck or WMAP. We have checked that in all our considerations, the DM $-$ nucleon cross-sections are beyond the present sensitivity of direct detection experiments such as XENON100 and LUX.

 In SM at $\Lambda_I \sim 10^{10}$~GeV, $\lambda$ vanish and becomes negative at higher energies. As $\beta_\lambda$ also becomes very small at high energies, it forbids $\lambda$ to attain a large negative value and thus protects the EW vacuum to go through a quick transition towards instability. As $\lambda\ra 0$ does not add to any known symmetry, we are not sure of the exact nature of new dynamics it is pointing to, or it could just be a fortuitous numerical coincidence. 
We observed that for our benchmark point, $\Lambda_I$ shifts by around an order of magnitude. So, the energy at which this unknown dynamics starts its action gets deferred in presence of the scalar $S$. Addition of $S$ does not help in realising the asymptotic safety scenario of gravity. But it is possible to have two degenerate vacua at somewhat lower energy, which depends on the parameter space under consideration.

In short, near-criticality of EW vacuum indicates presence of new dynamics other than the SM at a very high energy. In this work, we do not speculate about the nature such new physics, but introduce scalars at lower energies to demonstrate the influence of such scalars in shaping the minimum (if any) of the potential lying close to $\mpl$. In particular, we had shown that the onset scale of such high-energy dynamics may differ and the bounds on $M_H$ from stability considerations do change, in presence of these new scalars.

\vskip 20pt

\noindent{\bf Acknowledgments}\\
The work of NK is supported by a fellowship from UGC.
SR acknowledges support of seed grant from IIT Indore.
NK also acknowledges private communications with G.~Isidori,  A.~Strumia and M.F.~Zoller. SR is grateful to S.~Raychaudhuri for discussions.

\end{document}